\documentclass[twocolumn,showpacs,preprintnumbers,amsmath,amssymb]{revtex4}

\usepackage{graphicx}
\usepackage{dcolumn}
\usepackage{bm}
\usepackage{color}

\begin{document}

\title{Calculations for electron-impact excitation and ionization of beryllium}

\author{Oleg Zatsarinny$^1$}
\email[Electronic Address: ]{oleg.zatsarinny@drake.edu}
\author{Klaus Bartschat$^1$}
\email[Electronic Address: ]{klaus.bartschat@drake.edu}
\author{Dmitry V. Fursa$^2$}
\email[Electronic Address: ]{d.fursa@curtin.edu.au}
\author{Igor Bray$^2$}
\email[Electronic Address: ]{i.bray@curtin.edu.au}
\affiliation{$^1$Department of Physics and Astronomy, Drake University, Des Moines, Iowa, 50311, USA}
\affiliation{$^2$Curtin Institute for Computation and Department of Physics,
Astronomy and Medical Radiation Science, Curtin University, GPO Box U1987, Perth, WA 6845, Australia}

\date{\today}

\begin{abstract}
The \hbox{\emph{B}-spline} \hbox{\emph{R}-matrix} and the
convergent close-coupling methods are used to study electron collisions
with neutral beryllium over an energy range from threshold to 100~eV.
Coupling to the target continuum significantly affects the results for transitions from the ground state, but to a lesser extent
the strong transitions between excited states.
Cross sections are presented for selected transitions between low-lying physical bound states of beryllium,
as well as for elastic scattering, momentum transfer, and ionization.
The present cross sections for transitions from the ground state
from the two methods are in excellent agreement with each other, and
also with other available results based on non\-perturbative convergent pseudo\-state and time-dependent close-coupling models.
The elastic cross section at low energies is dominated by a prominent shape resonance.
The ionization from the $(2s2p)^3P$ and $(2s2p)^1P$ states strongly depends on the respective term.
The current predictions represent an extensive set of electron scattering data for neutral beryllium,
which should be sufficient for most modeling applications.
\end{abstract}

\pacs{34.80.Bm, 34.80.Dp}
\maketitle

\section{Introduction}
\label{sec:Introduction}

Beryllium is used as a surface material in the JET project~\cite{JETweb} and for the plasma-facing walls in ITER~\cite{ITERweb}.
This calls for accurate \hbox{e-Be} scattering data, as evidenced
by recent Coordinated Research Projects and Technical Meetings organized on this topic by the International Atomic Energy
Agency~\cite{IAEA-CRP,IAEA-Be-2015}.
Beryllium is among the most reactive elements, and its high chemical activity as well as its toxicity
make it virtually impossible to obtain reliable values of the electron-impact cross sections from direct measurements with
traditional setups.

Due to the lack of experimental data, researchers in plasma modeling currently have to rely
entirely on theore\-tical predictions. For this reason, it is important to estimate the accuracy
of the available theoretical data.
Extensive calculations utilizing state-of-the-art computational methods,
such as $R$-matrix with pseudo\-states (RMPS)~\cite{Bart1997,Bal2003}, convergent close-coupling (CCC)~\cite{Fursa1997},
and time-dependent close-coupling (TDCC)~\cite{Colgan2003}, were performed already more than a decade ago.
All these calculations indicate a slow convergence of the close-coupling expansion for certain transitions
and significant effects originating from coupling to the target continuum.
Due to the importance of the \hbox{e-Be} collision system, the topic remains under active investigation, with the
most recent CCC predictions published in 2015~\cite{BF2015}.

The purpose of the present paper is to provide an extensive and complete (for most modeling applications) set
of electron scattering data for neutral beryllium, including elastic scattering, momentum transfer,
excitation, and ionization from the ground state as well as a number of excited states,
including the metastable $(2s2p)^3P$ state, which is of particular importance for collisional radiative models.
The calculations reported below were carried out with the \hbox{$B$-spline}
\hbox{\hbox{$R$-matrix}} (close-coupling) code~\cite{BSR}. As an
independent check on these predictions, we also performed CCC
calculations containing a large number of pseudo\-states in the close-coupling expansion. Good agreement between these independent
calculations should provide additional confidence in the accuracy of
the obtained cross sections while any discrepancy would allow to
identify a  possible source of a problem.  

The distinct feature of the BSR approach is its ability to employ term-dependent non\-orthogonal
orbitals in the description of the target states. This allows us to optimize individual atomic
wave functions independently and thereby generate an accurate description of the target states
with relatively few configurations.
Over the past decade, the BSR code (along with its relativistic extension, DBSR~\cite{DBSR})
has been successfully applied to a number of targets, including those
with multiple open shells~\cite{rev}. Compared to some of those more complex systems,
neutral beryllium is relatively simple, in particular if only single-electron excitations from
the $2s^2$ filled outer sub\-shell are considered. For practical applications, this simplicity offers the advantage
of allowing for cross checks between the predictions from several
highly sophisticated methods, as done with  BSR and CCC calculations
presented here, rather than having to rely on 
just a single approach. 

In general, it is very helpful to investigate
electron-atom collision systems with different computational
approaches, particularly when high accuracy is required in
applications. Previous examples include, but are not limited to,  
e-Be$^+$~\cite{BB97jpbl}, e-Be$^{2+}$~\cite{MBB97jpbl}, and
e-Li~\cite{CPMGB01}. For quasi two-electron targets, as considered
here, the complexity of the electron-collision problem is such that 
substantial discrepancies may occur.  This was recently highlighted for the
e-Mg and e-Al$^+$ collision systems~\cite{BMF15}.

The manuscript is organized as follows: After discussing the description of the target
structure, we summarize the most important aspects of the collision calculations. This is
followed by a presentation of the cross sections for the most important transitions,
starting with elastic scattering from Be in its ground state.
Comparison of the results from the present BSR and CCC calculations with those from previous RMPS calculations, as well as
a systematic study of the effects of increasing channel coupling on the predictions,
provides a basis for estimating the accuracy of the present data\-set.

\section{Structure calculations}

\subsection{BSR}
The target states of beryllium in the present calculations were generated by combining
the multi-configuration Hartree-Fock (MCHF) and the \hbox{$B$-spline} box-based close-coupling
methods~\cite{BSR+}. We tried to account for the principal correlation effects, while
bearing in mind that the final multi-configuration expansions still
needed to be dealt with in the subsequent collision calculation with
one more electron to be coupled in. Since relativistic effects are small in beryllium,
certainly when it comes to their effect on electron collisions in practice,
we used the non\-relativistic $LS$-coupling approximation, with
the structure of the multi-channel target expansion chosen as
\begin{eqnarray}\label{eq:expn}
\Phi(2snl, LS)
  & = & \sum_{nl} a_{nl}^{LS}\left\{\phi(2s)P(nl)\right\}^{LS} \nonumber\\
  & + & \sum_{nl} b_{nl}^{LS} \left\{\phi(2p)P(nl)\right\}^{LS} \nonumber\\
  & + & \sum_{l,l'} c_{ll'}^{LS} \varphi(2l2l')^{LS}.
\label{eq:BSR-CI}
\end{eqnarray}
Here $P(nl)$ denotes the wave function of the outer valence electron,
while the $\phi$ and $\varphi$ functions stand for the configuration interaction (CI)
expansions of the corresponding ionic and specific atomic states, respectively.
These expansions  were generated in separate MCHF calculations
for each state using the MCHF program~\cite{MCHF}.
The expansion (\ref{eq:expn}) can be considered as a model for
the entire $2snl$ and $2pnl$ Rydberg series of the beryllium spectrum,
including auto\-ionizing states and continuum pseudo\-states.

The most correlated $2s^2$, $2s2p$ and $2p^2$ states
require more accurate descriptions. These states 
were represented by separate MCHF expansions, with the
orbitals specifically optimized for each given term.
Although abbreviated as $\varphi(2l2l')^{LS}$ in Eq.~(\ref{eq:BSR-CI}),
the particular expansions for these states include all single and double excitations
from the $2s$ and $2p$ orbitals to the $3l$ and $4l$ ($l = 0 - 3$)
correlated orbitals.

A more extensive description of core-valence correlation
would require additional ionic states by opening the
$1s^2$ shell. Their inclusion, however, would have considerably
increased the target expansions and made them no
longer tractable in the subsequent scattering calculations.

The unknown functions $P(nl)$ for the outer valence electron were expanded in a
\hbox{$B$-spline} basis, and the corresponding equations were solved subject to the
condition that the wave functions vanish at the boundary.
The \hbox{$B$-spline} coefficients for the valence orbitals $P(nl)$, along with the various expansion
coefficients in~(\ref{eq:expn}), were obtained by diagonalizing the $N$-electron atomic hamiltonian.
Since the \hbox{$B$-spline} bound-state close-coupling calculations generate different
non\-orthogonal sets of orbitals for each atomic state, their subsequent use is somewhat
complicated. On the other hand, our configuration expansions for the atomic target states
contained between 10 and 50 configurations for each state and hence could still be used in
the collision calculations with the available computational resources.

The number of spectroscopic bound states that can be generated in the above scheme depends on the
\hbox{$B$-spline} box radius. In the present calculations, the latter was set to $40\,a_0$,
where $a_0 = 0.529 \times 10^{-10}\,$m is the Bohr radius.
This allowed us to obtain accurate descriptions of the beryllium states with principal quantum number
for the valence electron up to $n=4$.

\subsection{CCC}
As for the BSR method, the nonrelativistic formulation of the CCC
method was adopted. We describe the Be atom by a model of two valence electrons above a
frozen Hartree-Fock (1s$^2$) core. This is the same approximation as
in the present BSR calculations.  The calculations of the Be  target states start
with the Hartree-Fock calculation for the Be$^+$ ion, which allows us to
obtain the $1s$ core orbital. The quasi one-electron Hartree-Fock hamiltonian
of the Be$^+$ ion is then diagonalized in a basis of Laguerre
functions. We add a one-electron polarization potential to the Be$^+$
hamiltonian to improve the agreement of the
Be$^+$ energy levels with the corresponding experimental values.
The number of Laguerre functions is $N_l=18-l$, and the
exponential fall-offs were chosen as $\lambda_l=0.9$ with the angular momentum ranging from $l=0$ to $l=3$.
The result of the diagonalization are one-electron (pseudo) states of
the Be$^+$ ion. In the present calculations we drop the two highest-lying
Be$^+$ states, as they lead to high-energy Be states that are
always closed at the energies considered in the present work.

The one-electron
basis is then used to expand the wave function of the Be atom in a set of
antisymmetric two-electron configurations. We include all configurations
of the type $\{2s,nl\}$ and $\{2p,nl\}$. In addition we include
$\{nl,n'l'\}$ configurations with $n, n' \le 3$. This choice of
configurations is practically the same as in~(\ref{eq:BSR-CI}). The latter
configurations allow us to accurately account for the electron-electron
correlations that are important for the ground state and the low-lying excited states of
the Be atom. The former set of configurations provides
a description of physical and pseudo excited states as well as a square-integrable
representation of the target continuum.
This set of configurations is used to diagonalize the quasi two-electron
hamiltonian of Be and thereby obtain the set of Be target states 
used in the scattering calculations. We include a two-electron
polarization potential to further improve the description of the ground state
and the low-lying excited states. These one- and
two-electron polarization potentials are employed  in the CCC calculations
but are not adopted in the BSR model. In practice, however, the
polarization potentials lead to only minor changes in
the target wave functions.

\begin{table}
\caption{\label{tab:E} Binding energies (in~eV) for the spectroscopic target states
included in the BSR and CCC  expansions.}
\begin{ruledtabular}
\begin{tabular}{llrrr}
 State    & Term    &  BSR     &  CCC      & NIST~\cite{NIST} \\
\hline
 $2s^2  $ & $^1S  $ & -9.287   & -9.312    & -9.323  \\
 $2s2p  $ & $^3P^o$ & -6.564   & -6.601    & -6.598  \\
 $2s2p  $ & $^1P^o$ & -4.000   & -3.952    & -4.045  \\
 $2s3s  $ & $^3S  $ & -2.854   & -2.860    & -2.865  \\
 $2s3s  $ & $^1S  $ & -2.518   & -2.539    & -2.544  \\
 $2p^2  $ & $^1D  $ & -2.240   & -2.238    & -2.270  \\
 $2s3p  $ & $^3P^o$ & -2.004   & -2.014    & -2.019  \\
 $2p^2  $ & $^3P  $ & -1.853   & -1.811    & -1.922  \\
 $2s3p  $ & $^1P^o$ & -1.836   & -1.838    & -1.860  \\
 $2s3d  $ & $^3D  $ & -1.623   & -1.625    & -1.629  \\
 $2s3d  $ & $^1D  $ & -1.314   & -1.314    & -1.335  \\
 $2s4s  $ & $^3S  $ & -1.321   & -1.324    & -1.325  \\
 $2s4s  $ & $^1S  $ & -1.222   & -1.230    & -1.233  \\
 $2s4p  $ & $^3P^o$ & -1.031   & -1.036    & -1.039  \\
 $2s4p  $ & $^1P^o$ & -0.999   & -1.004    & -1.011  \\
 $2s4d  $ & $^3D  $ & -0.896   & -0.895    & -0.899  \\
 $2s4f  $ & $^3F^o$ & -0.857   & -0.856    & -0.862  \\
 $2s4f  $ & $^1F^o$ & -0.857   & -0.856    & -0.862  \\
 $2s4d  $ & $^1D  $ & -0.779   & -0.777    & -0.795  \\
 $2s5s  $ & $^3S  $ & -0.765   & -0.742    & -0.767  \\
 $2s5s  $ & $^1S  $ & -0.723   & -0.690    & -0.728  \\
\end{tabular}
\end{ruledtabular}
\end{table}

\subsection{Energy Levels and Oscillator Strengths}
Table \ref{tab:E} compares the calculated spectrum of beryllium with the
values recommended by NIST~\cite{NIST} for various multiplets included in the scattering calculations described below.
Details of the evaluation procedure of the available data at the time of the original critical assessment
were given by Kramida and Martin~\cite{KM1997}, where the original sources can also be found. References to more
recent work, almost all theoretical, are also available from the NIST website.
The overall agreement between experiment and our theories is very satisfactory, in particular in light of the fact that
our structure descriptions are meant for the subsequent collision calculation, rather than as structure
models on their own. In the BSR model, specifically, the deviations from the NIST-recommended values in the energy splitting
are less than 45~meV for most states, with a few larger ones (up to 69~meV) seen only for the $2p^2$ states.
In the final BSR scattering calculations, we slightly adjusted the hamiltonian matrix elements, which
allowed us to use the experimental target energies.

\begin{table}
\caption{\label{tab:f} Comparison of oscillator strengths in Be: BSR and CCC models.}
\begin{ruledtabular}
\begin{tabular}{lllrlrr}
lower              &  upper            & \multicolumn{2}{c}{BSR}                  & \multicolumn{2}{c}{CCC}          & NIST       \\
level              &  level            & $ ~f_L $  & $\Delta$[\%]$^a$ & $ ~f_L $  & $\Delta[\%]^a$ & \cite{NIST} \\
\hline
  $ 2s^2\ ^1S   $  &  $ 2s2p\ ^1P^o $  & 1.36      &   4.3~   & 1.39      &   5.7~ &   1.37    \\
  $             $  &  $ 2s3p\ ^1P^o $  & 1.32E-02  &  10.1~   & 1.5E-02   &  27~   &   8.98E-3 \\
  $             $  &  $ 2s4p\ ^1P^o $  & 2.68E-04  &  20.8~   & 9E-6      &  -     &           \\
  $             $  &  $             $  &           &          &           &        &           \\
  $ 2s2p\ ^3P^o $  &  $ 2s3s\ ^3S   $  & 8.29E-02  &   4.4~   & 8.16E-02  &   6.8~ &   8.44E-2 \\
  $             $  &  $ 2p^2\ ^3P   $  & 4.55E-01  &   5.5~   & 4.57E-01  &   2.1~ &   4.47E-1 \\
  $             $  &  $ 2s3d\ ^3D   $  & 2.94E-01  &   1.3~   & 2.91E-01  &   1.0~ &   2.99E-1 \\
  $             $  &  $ 2s4s\ ^3S   $  & 1.17E-02  &   5.0~   & 1.15E-02  &   10.0~&   1.18E-2 \\
  $             $  &  $ 2s4d\ ^3D   $  & 1.01E-01  &   1.5~   & 9.76E-02  &   1.0~ &   9.61E-2 \\
  $             $  &  $             $  &           &          &           &        &           \\
  $ 2s2p\ ^1P^o $  &  $ 2s3s\ ^1S   $  & 1.18E-01  &   5.7~   & 1.22E-01  &   0.5~ &   1.15E-1 \\
  $             $  &  $ 2p^2\ ^1D   $  & 1.36E-03  &  12.0~   & 1.99E-03  &  55~   &           \\
  $             $  &  $ 2s3d\ ^1D   $  & 3.94E-01  &   1.3~   & 3.95E-01  &   2.0~ &   3.98E-1 \\
  $             $  &  $ 2s4s\ ^1S   $  & 8.53E-03  &   9.4~   & 8.85E-03  &   2.2~ &   9.81E-3 \\
  $             $  &  $ 2s4d\ ^1D   $  & 2.00E-01  &   0.5~   & 1.91E-01  &   1.3~ &   1.77E-1 \\
  $             $  &  $             $  &           &          &           &        &           \\
  $ 2s3s\ ^3S   $  &  $ 2s3p\ ^3P^o $  & 1.14      &   5.3~   & 1.13      &   3.0~ &   1.13    \\
  $             $  &  $ 2s4p\ ^3P^o $  & 4.40E-03  &  35.9~   & 3.36E-03  &  27.0~ &   3.51E-3 \\
  $             $  &  $             $  &           &          &           &        &           \\
  $ 2s3s\ ^1S   $  &  $ 2s3p\ ^1P^o $  & 9.79E-01  &   8.3~   & 9.73E-01  &   1.3~ &   9.57E-1 \\
  $             $  &  $ 2s4p\ ^1P^o $  & 8.53E-03  &  38.9~   & 7.82E-03  &  16.0~ &   9.68E-3 \\
  $             $  &  $             $  &           &          &           &        &           \\
  $ 2p^2\ ^1D   $  &  $ 2s3p\ ^1P^o $  & 7.13E-02  &   9.6~   & 7.09E-02  &   14~  &           \\
  $             $  &  $ 2s4p\ ^1P^o $  & 4.86E-03  &  12.1~   & 4.61E-03  &  25~   &           \\
  $             $  &  $ 2s4f\ ^1F^o $  & 1.61E-01  &   2.7~   & 1.60E-01  &   1.9~ &   1.56E-1 \\
  $             $  &  $             $  &           &          &           &        &           \\
  $ 2s3p\ ^3P^o $  &  $ 2s3d\ ^3D   $  & 4.98E-01  &   0.4~   & 5.04E-01  &   0.5~ &           \\
  $             $  &  $ 2s4s\ ^3S   $  & 2.19E-01  &   4.3~   & 2.16E-01  &   3.2~ &   2.18E-1 \\
  $             $  &  $ 2s4d\ ^3D   $  & 1.42E-01  &   0.4~   & 1.34E-01  &   0.2~ &   1.33E-1 \\
  $             $  &  $             $  &           &          &           &        &           \\
  $ 2s3p\ ^1P^o $  &  $ 2s3d\ ^1D   $  & 6.99E-01  &   3.7~   & 6.93E-01  &   3.2~ &   6.78E-1 \\
  $             $  &  $ 2s4s\ ^1S   $  & 2.11E-01  &   6.3~   & 2.14E-01  &   2.0~ &           \\
  $             $  &  $ 2s4d\ ^1D   $  & 1.92E-02  &  13.0~   & 1.74E-02  &  9.2~  &   1.74E-2 \\
  $             $  &  $             $  &           &          &           &        &           \\
  $ 2s3d\ ^3D   $  &  $ 2s4p\ ^3P^o $  & 8.17E-02  &   0.3~   & 8.25E-02  &   0.6~ &   8.09E-2 \\
  $             $  &  $ 2s4f\ ^3F^o $  & 8.90E-01  &   3.0~   & 8.75E-01  &   2.6~ &   8.74E-1 \\
  $             $  &  $             $  &           &          &           &        &           \\
  $ 2s4s\ ^3S   $  &  $ 2s4p\ ^3P^o $  & 1.63      &   2.8~   & 1.62      &   2.2~ &           \\
  $             $  &  $             $  &           &          &           &        &           \\
  $ 2s3d\ ^1D   $  &  $ 2s4p\ ^1P^o $  & 2.06      &   4.6~   & 2.06      &   5.0~ &           \\
  $             $  &  $ 2s4f\ ^1F^o $  & 1.03      &   2.9~   & 1.02      &   3.4~ &   1.02    \\
  $             $  &  $             $  &           &          &           &        &           \\
  $ 2s4s\ ^1S   $  &  $ 2s4p\ ^1P^o $  & 1.45      &   5.0~   & 1.45      &   2.0~ &           \\
  $             $  &  $             $  &           &          &           &        &           \\
  $ 2s4p\ ^3P^o $  &  $ 2s4d\ ^3D   $  & 7.88E-01  &   0.3~   & 8.08E-01  &   0.0~ &           \\
  $             $  &  $             $  &           &          &           &        &           \\
  $ 2s4p\ ^1P^o $  &  $ 2s4d\ ^1D   $  & 1.21      &   2.4~   & 1.21      &   1.7~ &           \\
  $             $  &  $             $  &           &          &           &        &           \\
  $ 2s4d\ ^3D   $  &  $ 2s4f\ ^3F^o $  & 1.33E-01  &   4.6~   & 1.46E-01  &   3.5~ &           \\
  $             $  &  $             $  &           &          &           &        &           \\
  $ 2s4f\ ^1F^o $  &  $ 2s4d\ ^1D   $  & 1.51E-01  &   6.3~   & 1.46E-01  &   7.3~ &           \\
\end{tabular}
\end{ruledtabular}
$^a$ Percentage difference between the $f_L$ and $f_V$ values.  \hfill     \\
\end{table}

In the CCC model, the exponential fall-offs of the Laguerre basis were chosen in
such a way that the target states of Be  are well described up to
principal quantum number $n=4$. These states were referred to as
the spectroscopic bound states in the BSR structure calculations above. 
As seen from Table~\ref{tab:E}, the accuracy of the CCC target states is of
the same level as for the target states obtained in the BSR
calculations. This is particularly pleasing given the different underlying
one-electron basis orbitals in the two methods.

The quality of our target descriptions can be further assessed by comparing the
results for the oscillator strengths of various transitions with experimental data and other
theoretical predictions. Accurate oscillator strengths
are very important to obtain reliable absolute values for the excitation cross sections,
especially for optically allowed transitions at high incident electron energies.
A comparison of our oscillator strengths is given in \hbox{Table~\ref{tab:f}} with
the recommended data from the NIST compilation~\cite{NIST}.
The latter recommendations are based on the semi\-relativistic Breit-Pauli calculations by Tachiev and Froese Fischer~\cite{TFF1999}.
The $f$-values for the fine-structure transitions were converted to the corresponding multiplet $LS$-values.
We see good agreement for all these transitions.
Table~\ref{tab:f} also contains the ratio of theoretical
oscillator strengths obtained in the length and velocity forms of the electric dipole operator.
This ratio can, to some extent, be considered an accuracy indicator for the calculated \emph{f}-values.
For most transitions, the length ($f_L$) and velocity ($f_V$) values agree within a few percent, and
once again the quality of the BSR and CCC results is comparable.

\section{Collision calculations}

\subsection{BSR} 
The close-coupling expansion in our most extensive model contained 660 states of neutral beryllium,
with 29 states representing the bound spectrum and the remaining 631 the target continuum corresponding to ionization of
the $2s^2$ subshell. We included all singlet and quartet target states
of configurations $2snl$ and $2pnl$ with orbital angular momentum
$l=0-3$ for the outer electron and total orbital angular momenta $L=0-4$.
The continuum pseudo\-states in the present calculations cover the energy region up to 60~eV above the ionization limit.
This model will be referred to as \hbox{BSR-660} below.  Other time-independent close-coupling
models will be labeled similarly by indicating the approach (RMPS, CCC, BSR), followed by the number of states
included in the expansion.

The close-coupling equations were solved by means of the \hbox{\hbox{$R$-matrix}} method, using a
parallelized version of the BSR code~\cite{BSR}.
The distinctive feature of the method is the use
of \hbox{$B$-spline}s as a universal basis to represent the scattering
orbitals in the inner region of $r \le a$. Hence, the
\hbox{\hbox{$R$-matrix} expansion} in this region takes the form
\begin{eqnarray}\label{eq:RM}
\Psi_k(x_1,\ldots,x_{N+1})  = ~~~~~~~~~~~~~~~~~~~~~~~~~~~~~~~~~~~~~~~~~~& \nonumber\\
~~~{\cal A} \sum_{ij} \bar{\Phi}_i(x_1,\ldots,x_N;\hat{\mathbf{r}}_{N+1}\sigma_{N+1})
\,r_{N+1}^{-1}\,B_j(r_{N+1})\,a_{ijk} \!\!\!\!\!\!\! & \nonumber\\
 +  \sum_i \chi_i(x_1,\ldots,x_{N+1})\,b_{ik}.~~~~~~~
\end{eqnarray}
Here the $\bar{\Phi}_{i}$ denote the channel functions constructed from the
$N$-electron target states, while the splines $B_{j}(r)$ represent the continuum orbitals.
The $\chi_i$ are additional $(N\!+\!1)$-electron bound states.
In standard \hbox{\hbox{$R$-matrix}} calculations~\cite{Burke}, the latter are included
one configuration at a time to ensure completeness of the total
trial wave function and to compensate for orthogonality constraints imposed on
the continuum orbitals.
The use of non\-orthogonal one-electron radial functions in the BSR method, on the other hand,
allows us to completely avoid these configurations for compensating orthogonality restrictions.
Sometimes, explicit bound channels in BSR calculations are used for a more accurate description
of the true bound states in the collision system.
In the present calculations, however, we did not employ any $(N+1)$-electron correlation
configurations in the expansion~(\ref{eq:RM}).

The \hbox{$B$-spline} basis in the present calculations contained 80~splines of order~8, with
the maximum interval in this grid of $0.65\,a_0$. This is sufficient
for a good representation of the scattering electron wave functions for energies up to 150~eV.
The \hbox{BSR-660} collision model contained up to 1,566~scattering channels,
leading to generalized eigenvalue problems with matrix dimensions up to 100,000 in the \hbox{\hbox{$B$-spline}} basis.
Explicit numerical calculations were performed for partial waves with total orbital angular momenta $L \le 25$.
Taking into account the total spin and parity leads to 104~partial waves overall.
A top-up procedure based on the geometric-series approximation was used to estimate the contribution
from higher $L$~values if needed. The calculation for the external region
was performed using a parallelized version of the STGF program~\cite{stgf}.

\subsection{CCC} 
All states obtained from the diagonalization of the Be hamiltonian were
included in the close-coupling expansion of the total wave function.
Since the total number of target states is 409, the corresponding
calculations will be referred to as CCC-409. This model includes
singlet and triplet states of positive and negative parity with total
orbital angular momentum $L = 0 - 4$. Similarly to the BSR-660 model, the
number of negative energy states is 29 and the remaining states are
positive energy pseudo\-states modeling the target continuum. The 
positive-energy pseudo\-states span the energy region up to 130~eV for \hbox{$S$-states},
120~eV for \hbox{$P$-states}, 100~eV for \hbox{$D$-states}, 80~eV for \hbox{$F$-states}, 
and 70~eV for \hbox{$G$-states}. While the negative-energy states in
the BSR-660 and \hbox{CCC-409} models are practically the same, the distribution of the
positive energy pseudo\-states is very different. The use of the
Laguerre basis  in the CCC method leads to an exponential distribution
of the states with energy that has a larger density of states at small
energies. The $B$-spline basis, on the other hand, leads to a uniform
distribution of the states with energy. This requires a larger
number of states to cover approximately the same energy region.

In the CCC method the close-coupling expansion of the total wave function 
is inserted into the Schr\"{o}dinger equation, which is
transformed into momentum space, where it results in a set of coupled Lippmann-Schwinger
equations for the $T$-matrix that is solved by standard
techniques.  The largest number of channels in the CCC-409
calculations was about 1,000 and the solution method required solving a set of
linear equations of dimension up to 90,000 for each incident electron energy. The solution was
obtained via a massively parallel hybrid OpenMP-MPI implementation
that is standard for the entire suite of CCC programs (including
the relativistic~\cite{FB08l} and molecular \cite{ZetalH2+} formulations). The
calculations in the \hbox{CCC-409} model were performed for partial waves up to the total
orbital angular momentum $L=15$. We then used an analytical Born subtraction
technique to account for larger partial waves (formally up to
infinity). To verify the accuracy of the analytical Born-subtraction
technique, we performed calculations up to $L=25$ at a number
of energies and found negligible variations.

\begin{figure}
\includegraphics[width=0.45\textwidth]{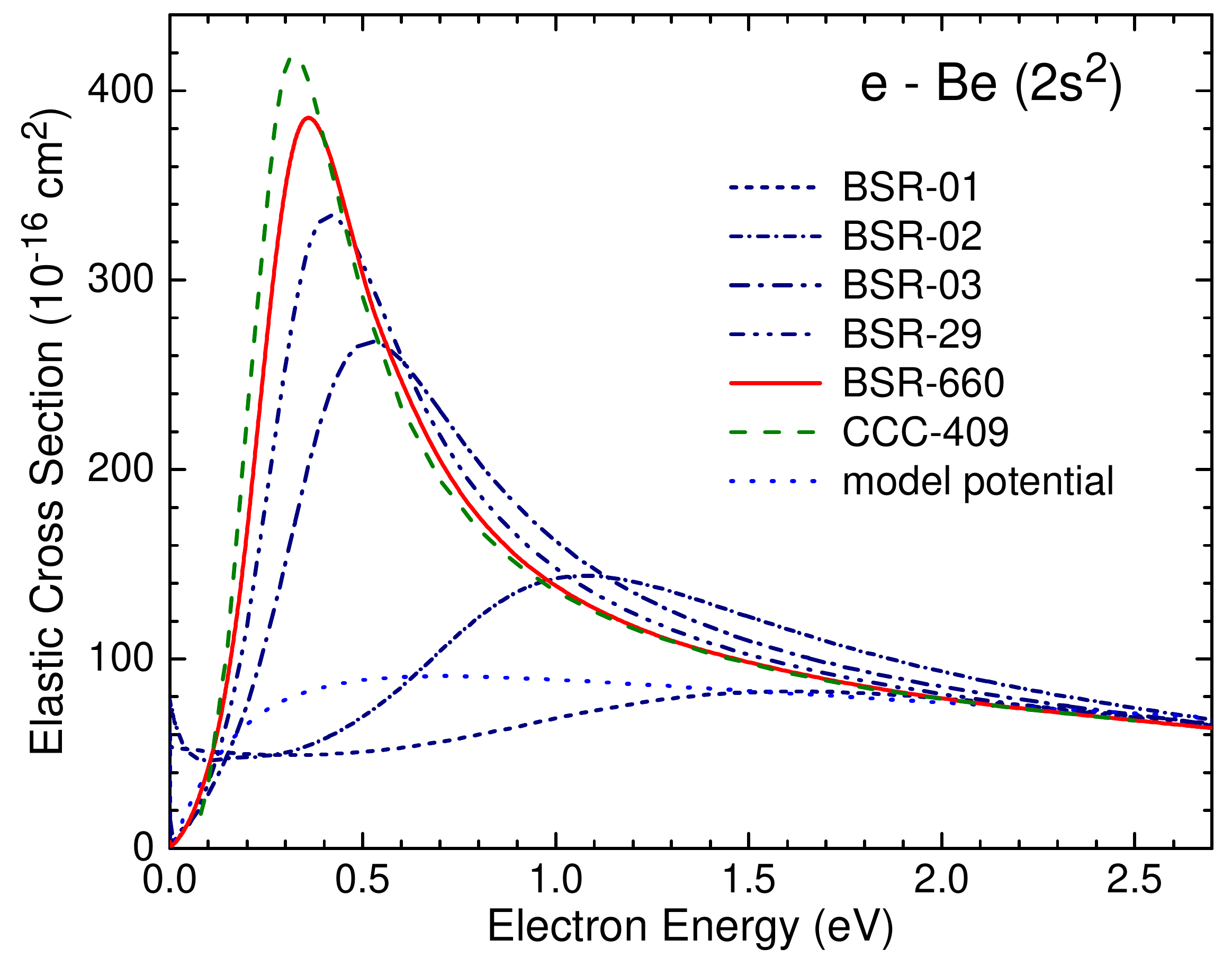}
\caption{\label{fig:elas} (Color online) Cross sections for elastic
electron scattering from beryllium atoms in their $(2s^2)^1S$ ground state at low energies
in the region of the shape resonance.
We present several BSR calculations to illustrate the convergence pattern.
Also shown are the model-potential calculations by Reid and Wadehra~\cite{Reid2014}.}
\end{figure}

\begin{figure}
\includegraphics[width=0.45\textwidth]{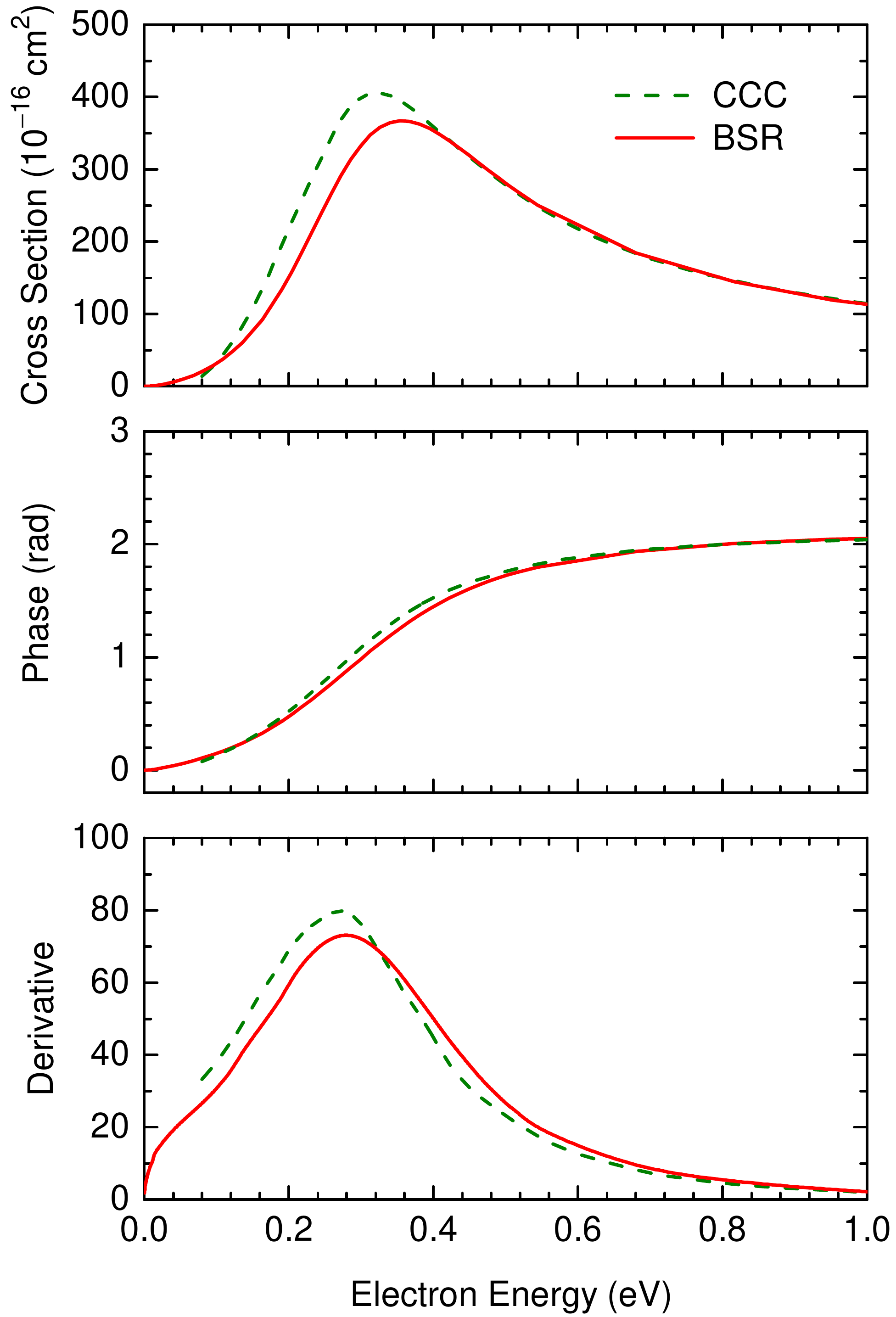}
\caption{\label{fig:phase} (Color online) Cross sections, phase shift, and its derivative
for the $P$ partial wave for elastic electron scattering from beryllium atoms
in the region of the shape resonance, as obtained in the \hbox{BSR-600} and \hbox{CCC-409} models.}
\end{figure}

\section{Results and Discussion}

\subsection{Elastic Scattering}

Results for the elastic cross section for electron scattering from
the beryllium in its ground state are presented in Fig.~\ref{fig:elas},
where we compare the predictions from several BSR calculations to
illustrate the convergence with the number of states included in the
close-coupling expansion.  As seen in the figure, the low-energy
regime is dominated by a shape resonance, for which the convergence of
the theoretical predictions is very slow.  This is important but
ultimately not surprising, since the very same effect was seen in
\hbox{e-Mg} scattering~\cite{Mg2004}. As expected, the \hbox{CCC-409} predictions are in
good agreement with those from the \hbox{BSR-660} model.

The above shape resonance in elastic e-Be scattering has been the subject of numerous calculations
with different methods. An overview of the many predictions
is given in Table~III of~\cite{Tsednee2015}. The results differ considerably,
ranging for positions from 0.1~eV to 1.2~eV and widths
from 0.14~eV to 1.78~eV, respectively. In this respect, it is
worth\-while to provide the resonance parameters from direct scattering calculations.

The standard determination of such resonance parameters from collision calculations
is based on the analysis of the
phase shift in the corresponding partial wave. In the vicinity of a resonance, the phase shift $\delta$
behaves as
\begin{eqnarray}\label{eq:phase}
\delta(E) = \delta^0(E) + \tan^{-1}\frac{\Gamma/2}{E_r-E}.
\end{eqnarray}
Assuming that the background phase shift~$\delta^0(E)$ is a smooth function of energy,
the resonance width~$\Gamma$ is determined from the inverse of the energy derivative of the
phase shift $\delta$ at the resonance energy $E_r$ via
\begin{eqnarray}\label{eq:gamma}
\Gamma = 2\,\left[\frac{d\delta}{dE}\right]_{E=E_r}^{-1}.
\end{eqnarray}
Such an analysis for the $P$ partial wave is given in Fig.~\ref{fig:phase},
and the corresponding resonance parameters are listed in Table~\ref{tab:G}.

\begin{table}[h!]
\caption{\label{tab:G} Position (E) and width ($\Gamma$) of the shape resonance (in~eV).}
\begin{ruledtabular}
\begin{tabular}{ccccc}
           & \multicolumn{2}{c}{cross section maximum}   &  \multicolumn{2}{c}{phase analysis}\\   
 Method    & $E_r$ & $\Gamma$  &  $E_r$ & $\Gamma$      \\
\hline
 BSR       & 0.354 & 0.461   & 0.284    & 0.372  \\
 CCC       & 0.320 & 0.434   & 0.269    & 0.341  \\
\end{tabular}
\end{ruledtabular}
\end{table}

We note, however, that this procedure is somewhat ambiguous in the present case.
Since the resonance is very wide and located close to the elastic threshold, 
the energy dependence of the phase shift given in Eq.~(\ref{eq:phase}) is disturbed. As a result, the phase
shift increases by less than $\pi$ radians as the energy passes through the resonance.

Another possibility, although not unique either, is to define the resonance parameters from the
analysis of the relevant partial-wave (here the \hbox{$P$-wave}) cross section.
An estimate for the resonance energy is then obtained from the maximum 
of the cross section while the (full) width is determined from half the height of this maximum.
Table~\ref{tab:G} also presents the resonance parameters generated in this way.  
The difference between the BSR and CCC predictions, and the difference between the values obtained in
the two ways of analyzing the results,
provide an indication of the likely uncertainty of the resonance parameters in the present calculations.
Taking the averages of the results obtained in the schemes outlined above, we estimate the
position at about 0.31~eV $\pm~0.04$~eV above the elastic threshold with a width of 0.40~eV $\pm~0.06$~eV.
These parameters differ considerably from the numerous
results obtained with model potentials~\cite{Reid2014}
or complex-rotation-based methods~\cite{Tsednee2015} methods.

\subsection{Excitation}

\begin{figure}[b!]
\includegraphics[width=0.48\textwidth]{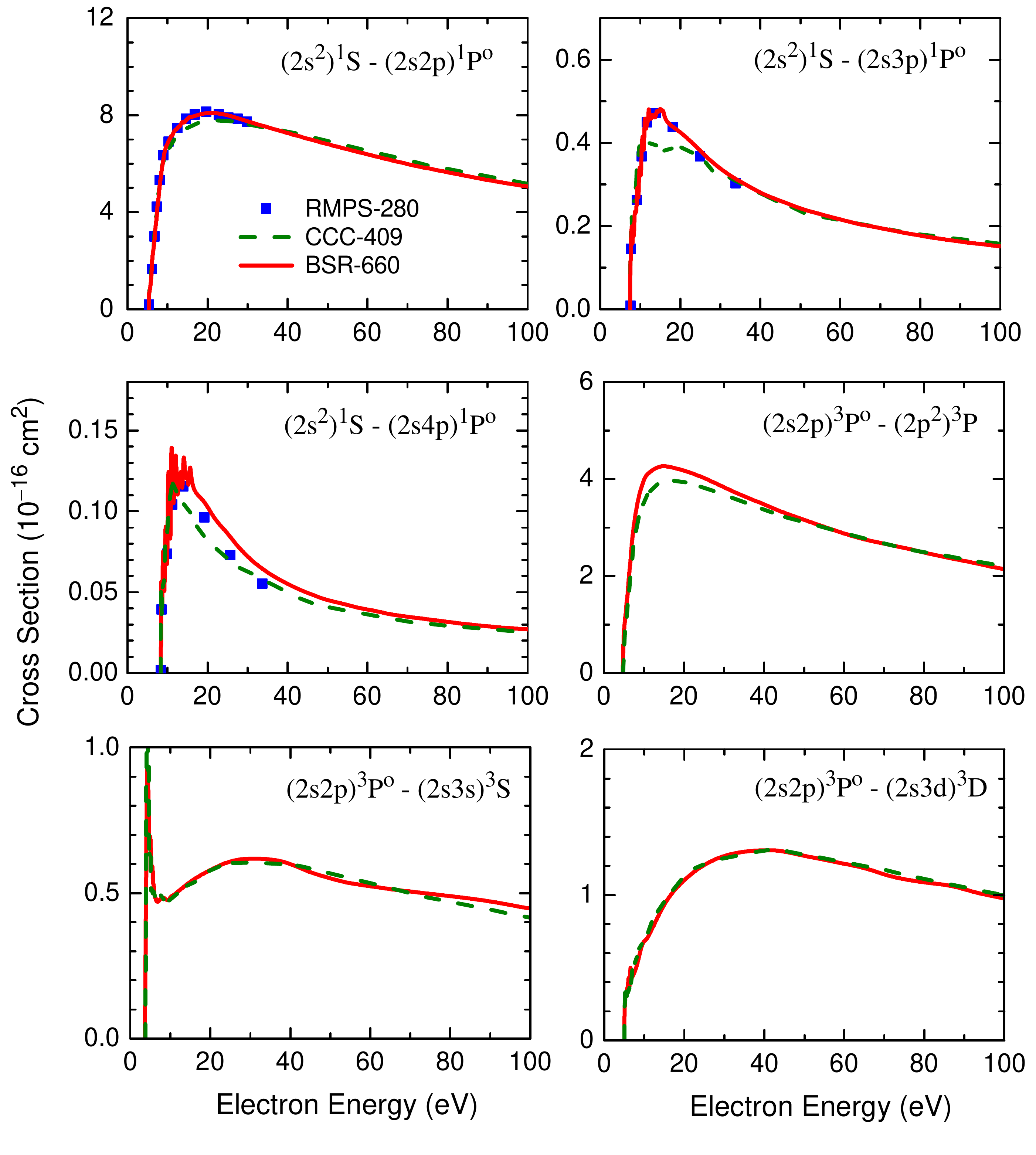}
\caption{\label{fig:dip} (Color online) Cross sections as a function of collision energy for
selected dipole transitions in beryllium. The present \hbox{BSR-660} and \hbox{CCC-409} results are compared
with those from an earlier \hbox{RMPS-280}~\cite{Bal2003} calculation.}
\end{figure}

\begin{figure}
\includegraphics[width=0.48\textwidth]{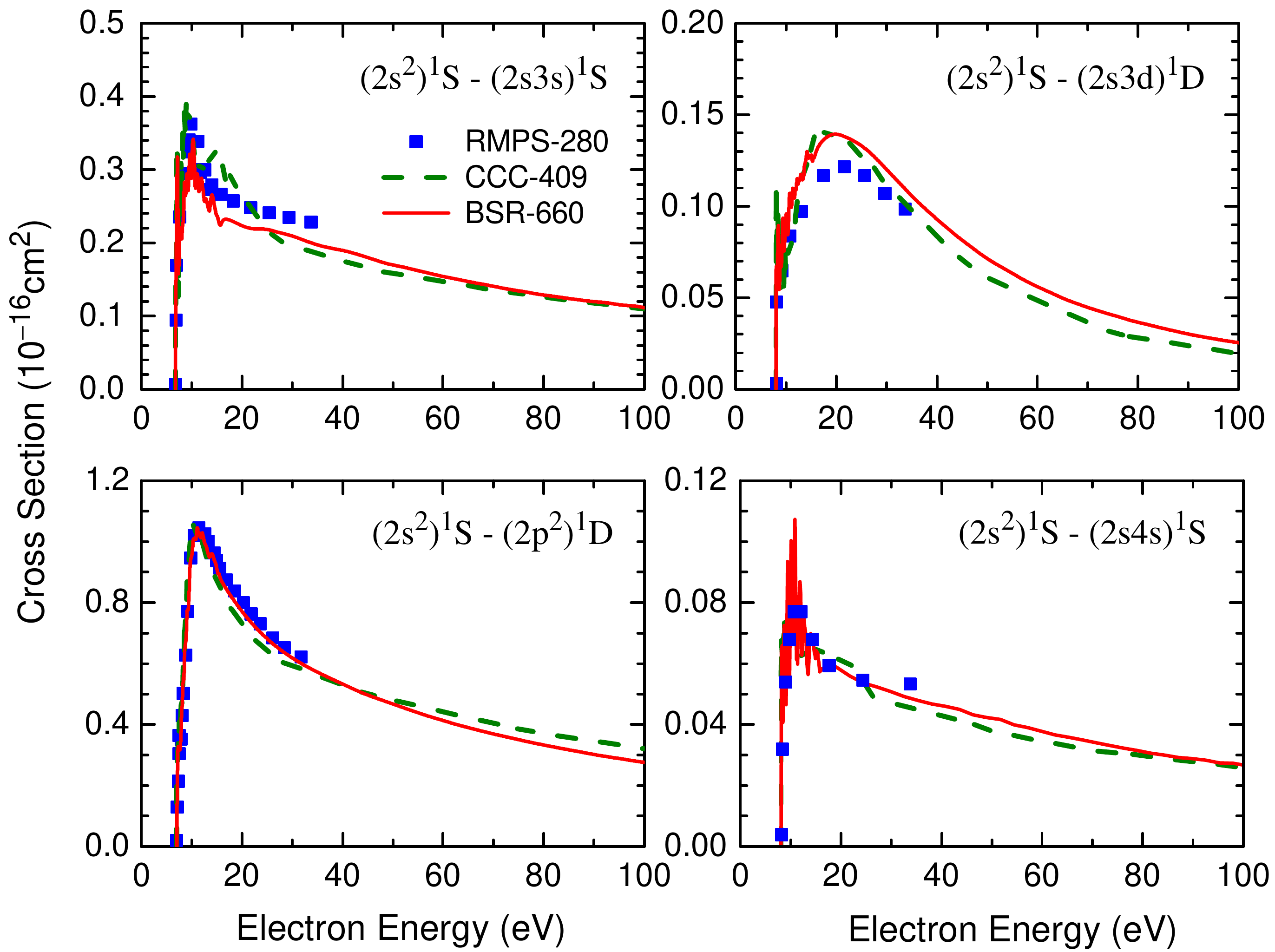}
\caption{\label{fig:ndip} (Color online) Cross sections as a function of collision energy for selected
non\-dipole transitions in beryllium. The present \hbox{BSR-660} and \hbox{CCC-409} results are compared
with those from an earlier \hbox{RMPS-280}~\cite{Bal2003} calculation.}
\end{figure}

\begin{figure}
\includegraphics[width=0.48\textwidth]{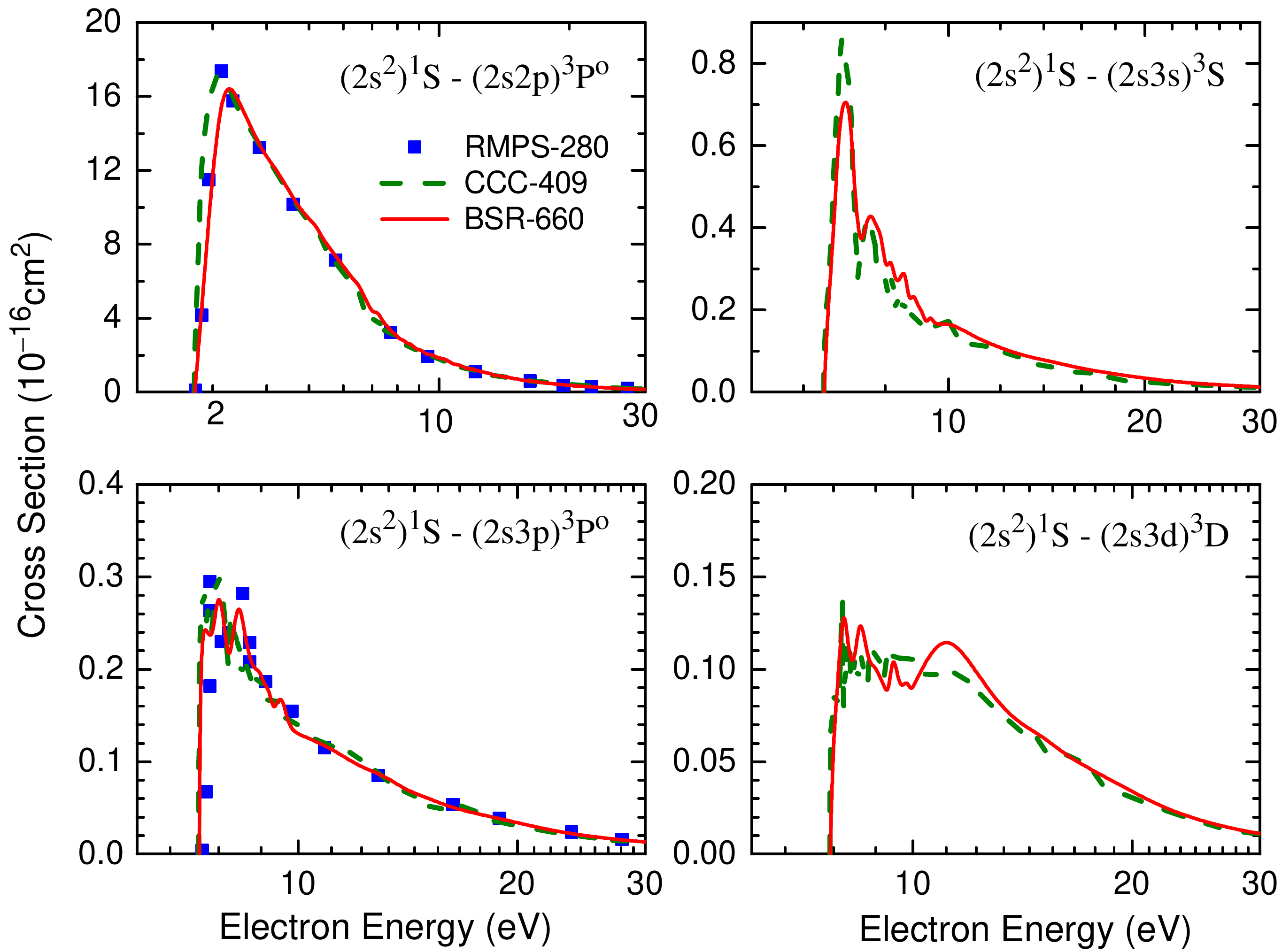}
\caption{\label{fig:exch} (Color online) Cross sections as a function of collision energy for selected
exchange transitions in beryllium. The present \hbox{BSR-660} and \hbox{CCC-409} results are compared
with those from an earlier \hbox{RMPS-280}~\cite{Bal2003} calculation.}
\end{figure}

Cross sections as a function of energy for the most important transitions from the $(2s^2)^1S$ ground state
and the metastable $(2s2p)^3P^o$ excited state are presented in Figs.~\ref{fig:dip}$-$\ref{fig:exch}
for dipole, non\-dipole, and exchange transitions, respectively.
We compare our BSR and CCC  predictions with the published RMPS~\cite{Bal2003} results.
For the very weak transitions, we notice some resonance-like structure near and slightly above the
ionization threshold.  These structures are, indeed, typical for pseudo\-state calculations, even if
the $N$-electron and $(N\!+\!1)$-electron configurations are constructed in a fully consistent manner
with each other. The degree of
visibility depends on the number of points displayed.  Note, however, that rate coefficients
involve convolution of the cross sections with the appropriate electron energy distribution function.
This, together with the small values of the cross sections for which these structures appear,
should ensure that there are no serious
problems in collisional radiative models that employ our results.
Overall, we trust that the very close agreement between several independently obtained results further
solidifies the confidence of the plasma modeling community in using these data\-sets.

\subsection{Ionization and Grand Total Cross Section}

\begin{figure}
\includegraphics[width=0.45\textwidth]{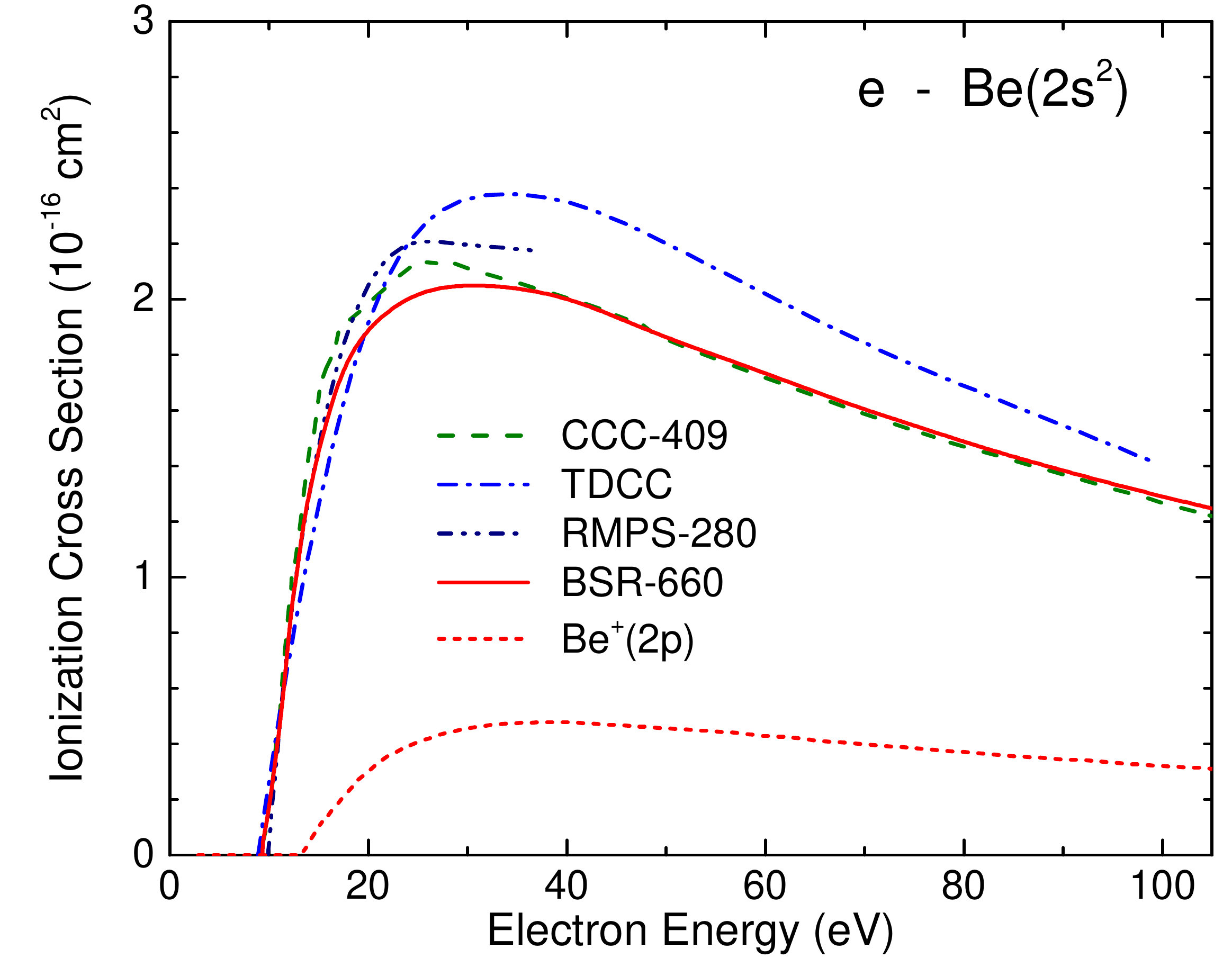}
\caption{\label{fig:ion} (Color online) Cross section for
electron-impact ionization of beryllium from the $(2s^2)^1S$ ground state.
The present \hbox{BSR-660} and \hbox{CCC-409} results are compared with those from earlier \hbox{RMPS-280}~\cite{Bal2003}
and TDCC~\cite{Colgan2003} calculations.
Also shown is the partial cross section for producing the excited $1s^22p$ state of Be$^+$ (obtained with \hbox{BSR-660}).}
\end{figure}

\begin{figure}
\includegraphics[width=0.45\textwidth]{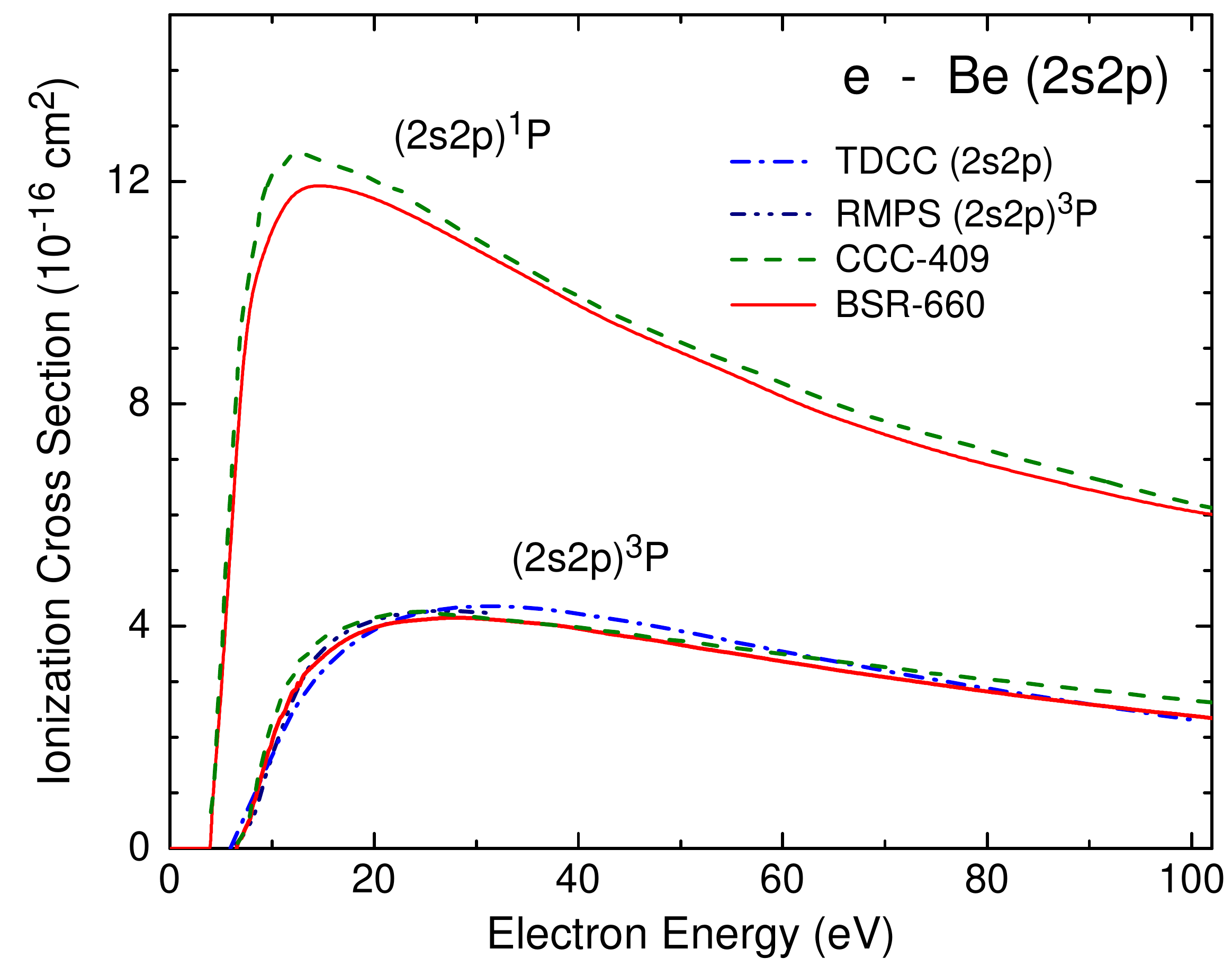}
\caption{\label{fig:ion2} (Color online) Electron-impact ionization cross sections for neutral beryllium
from the first excited $2s2p$ configuration.
The present \hbox{BSR-660} and \hbox{CCC-409} results are compared with those from earlier \hbox{RMPS-280}~\cite{Bal2003}
and TDCC~\cite{Colgan2003} calculations.}
\end{figure}

Ionization cross sections are presented in Figs.~\ref{fig:ion}
and~\ref{fig:ion2}. The \hbox{BSR-660} and \hbox{CCC-409} ionization cross sections
were obtained as the sum of the excitation cross section to all beryllium auto\-ionizing states and the continuum pseudo\-states.
We assumed that the radiative decay of the auto\-ionizing states is negligible in comparison to the auto\-ionization channel.
We find very good agreement between the present \hbox{BSR-660} and CCC-409 results,
but the agreement with the earlier RMPS~\cite{Bal2003} and TDCC results~\cite{Colgan2003} is also
very satisfactory for ionization from both the $(2s^2)^1S$ ground state (see Fig.~\ref{fig:ion}) and the metastable excited
$(2s2p)^3P^o$ state (Fig.~\ref{fig:ion2}).

Figure~\ref{fig:ion2} reveals a strong term dependence in ionization of the $(2s2p)^3P^o$ and $^1P^o$ states.
This is essentially due to the well-known term dependence of the $2p$ orbital~\cite{CFF}.  Since the TDCC model employed a
$2p$ orbital that is close to the Hartree-Fock orbital optimized on the $(2s2p)^3P^o$ state, the TDCC results displayed here are expected
to be most appropriate for this state.

\begin{figure}
\includegraphics[width=0.45\textwidth]{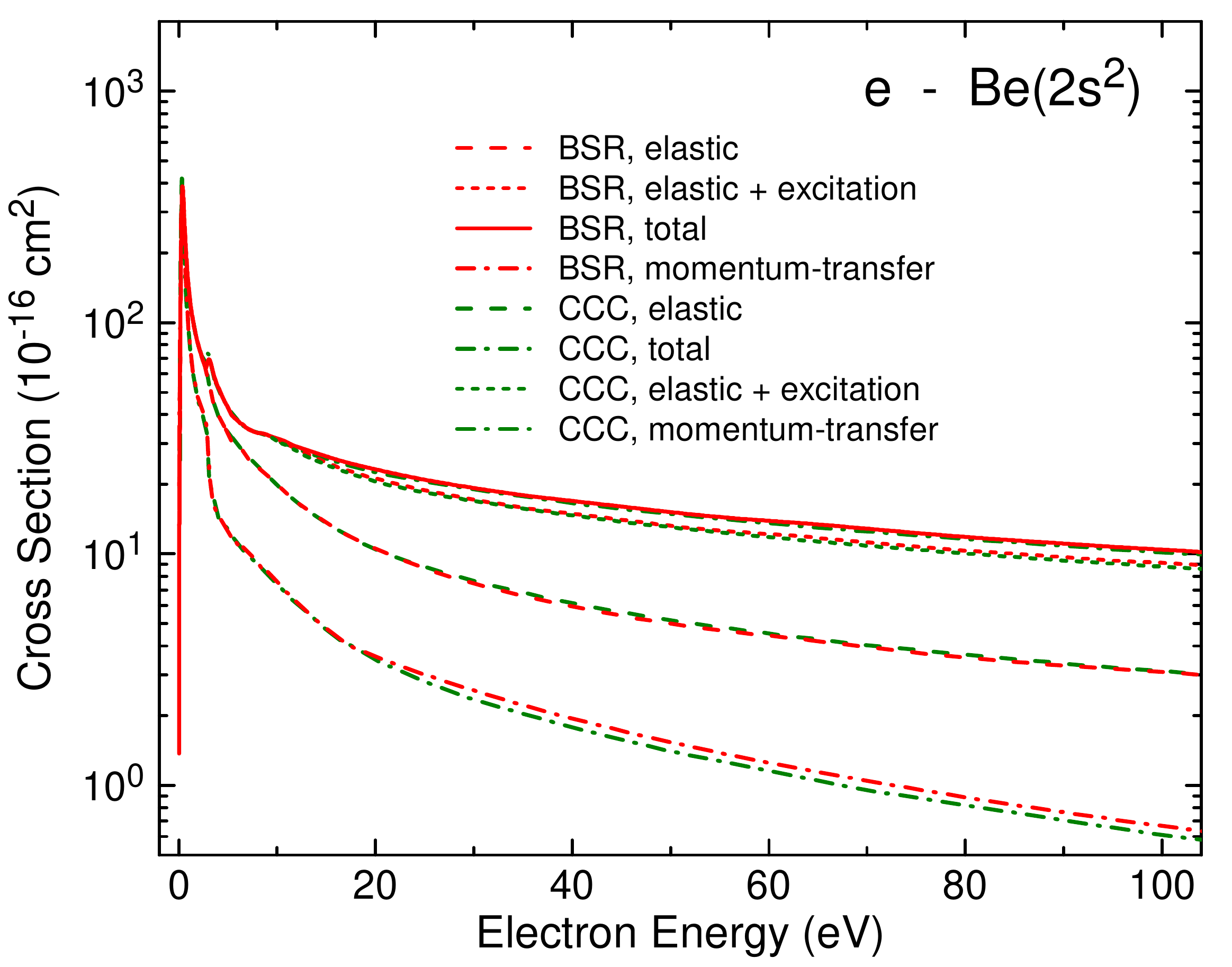}
\caption{\label{fig:total} (Color online) \hbox{BSR-660} and \hbox{CCC-409} grand total cross section
for electron collisions with beryllium atoms in their $(2s^2)^1S$ ground state, along
with the contributions from elastic scattering alone as well as elastic scattering plus excitation
processes.  Also shown is the momentum-transfer cross section.}
\end{figure}

Finally, Fig.~\ref{fig:total} exhibits the grand total cross
section for electron collisions with beryllium atoms in their $(2s^2)^1S$ ground state,
i.e., the sum of angle-integrated elastic, excitation, and ionization cross sections.
While the elastic cross section provides the largest contribution over the
energy range shown, excitation also contributes substantially, approaching 50\% for incident energies above 50~eV.
Overall, ionization processes represent less than 10\% of the grand total cross section.
Since the momentum-transfer rather than the elastic cross section
is typically important for plasma modeling, it is also shown in Fig.~\ref{fig:total}.

\section{Summary}

We have presented an extensive set of electron scattering data for
neutral beryllium, including elastic scattering, momentum transfer,
excitation, and ionization processes.  While state-to-state excitation
cross sections were obtained for all transitions between the lowest 21
states of beryllium, results were presented for only a few selected
transitions.  The calculations were performed with a parallel version
of the BSR code~\cite{BSR}, in which a \hbox{\emph{B}-spline} basis is
employed to represent the continuum functions inside the
\hbox{\hbox{$R$-matrix}} sphere.  Furthermore, we utilize
non\-orthogonal orbitals, both in constructing the target states and
in representing the scattering functions. In order to independently
verify the accuracy of the BSR calculations, we also carried out
CCC calculations with an entirely different formulation of the problem and
the associated computer code. Very good agreement between
the BSR and CCC results was found for all calculated cross sections.

The present calculations were motivated to a large extent by the importance of accurate and
thoroughly assessed \hbox{e-Be}
collision data. For excitation as well as ionization from the ground state and the most important metastable
$(2s2p)^3P$ state, we essentially confirm, where available, results from earlier RMPS~\cite{Bal2003} and TDCC~\cite{Colgan2003} calculations.  
Furthermore, we found a significant term dependence in the ionization results for the $(2s2p)^3P$ and $(2s2p)^1P$ states, respectively.

The elastic cross section at low energies is dominated by a strong shape resonance in the $L=1$, odd-parity channel.
Since this resonance is likely of critical importance for transport processes,
we carried out a systematic convergence study for its parameters.  The present results, namely a
position of about 0.31~eV $\pm~0.04$~eV above the elastic threshold with a width of 0.40~eV $\pm~0.06$~eV, are very different from
recent predictions based on a model-potential method~\cite{Reid2014} and also on a
complex-rotation approach~\cite{Tsednee2015}.

Based on our convergence studies, as well as a detailed comparison between the BSR and CCC results,
we estimate the accuracy of the present dataset to be 10\% or better when these data are used
to obtain the relevant rate coefficients for plasma modeling applications.
Electronic files with the current results, for electron energies up to 100~eV,
are available from the authors upon request.

\section*{Acknowledgments}
We thank Drs.\ B.~J.~Braams and H.-K.~Chung for drawing our attention to the
continued importance of accurate data for the e-Be collision problem, and  
Drs.\ \hbox{J.~Colgan} and \hbox{C.~P.~Ballance} for clarifying comments
on the manuscript.  The work of OZ and KB was supported by the United States
National Science Foundation under grants No.~PHY-1212450,
No.~PHY-1403245, and No.~PHY-1520970, and by the super\-computer
allocation No.~PHY-090031 within the eXtreme Science and Engineering Discovery Environment. 
The BSR calculations were carried out on
Stampede at the Texas Advanced Computing Center and on Gordon at the
San Diego Supercomputer Center.  DVF and IB acknowledge support
from Curtin University and resources provided by the Pawsey
Supercomputing Centre with funding from the Australian Government
and the Government of Western Australia.

\end{document}